\documentstyle[12pt]{article}
\newcommand{\be}{\begin{equation}}
\newcommand{\ee}{\end{equation}}
\begin{document}
\vspace*{0.5 in}
\begin{center}
\Large
{\bf Scaling of Heteroepitaxial Island Sizes}
\end{center}
\bigskip \bigskip
\large
\centerline{C. Ratsch and A. Zangwill}
\centerline{\it School of Physics,
Georgia Institute of Technology,
Atlanta, GA 30332}
\bigskip
\centerline{P.~\v{S}milauer$^{\dagger}$}
\centerline{\it Interdisciplinary Research Centre for Semiconductor
Materials}
\centerline{\it Imperial College, London SW7 2BZ,
United Kingdom}
\bigskip\bigskip \bigskip \bigskip

\centerline{\bf Abstract}
\medskip

\noindent
Monte Carlo simulations
of an atomistic solid-on-solid model are used
to study the effect of lattice misfit on the
distribution of two-dimensional
islands sizes as a function of coverage $\Theta$
in the submonolayer aggregation regime of epitaxial growth.
Misfit promotes the detachment of atoms from the perimeter  of
large pseudomorphic islands and thus favors their dissolution into smaller
islands that relieve strain more efficiently.
The number density of islands composed
of $s$ atoms exhibits scaling in the form
\mbox{$N_s(\Theta) \sim \Theta / \langle s \rangle^2 \, g(s/\langle s
\rangle$)}
where $\langle s \rangle$ is the average island size.  Unlike the case
of homoepitaxy, a  rate equation theory based on this observation
leads to qualitatively different
behavior than observed in the simulations.
\bigskip


\vfill
\noindent{\normalsize
$^{\dag}$ Also at The Blackett Laboratory, Imperial College, London SW7 2BZ,
United Kingdom;
on leave from
the
Institute of Physics, Czech Academy of Science,
Cukrovarnick\'{a} 10, 16200 Praha 6, Czech Republic.}

\newpage

The morphology that obtains when atoms of one material are deposited onto
a substrate of a dissimilar material is a central concern in current
efforts to
fabricate nanostructures {\it in situ} during growth.
Equilibrium considerations make clear that small, undislocated,
three-dimensional
islands
achieve significant epitaxial strain relief by lattice relaxation at the island
edges \cite{Ratsch}. But such relaxation occurs at the edges of
two-dimensional (2D) heteroepitaxial islands as well \cite{Massies}. Thus,
since there is evidence that 2D islands form a template from which
3D island structures evolve \cite{Joe},
it seems appropriate to focus attention on
some of the kinetic aspects of
heteroepitaxy even
before a full monolayer has been deposited.
To this end,
we study here the evolution of 2D island size
distributions
with a simple simulation
model of epitaxial
growth suited to
the case when the difference between the
deposited material and the substrate is completely characterized by their
lattice misfit.

The computations reported here
generalize a previously successful Monte Carlo model of homoepitaxy \cite{SOS}.
There,
atoms are
deposited at random onto the (001) surface sites of a simple cubic
lattice (with unit lattice
constant) at an average rate $F$. No vacancies or overhangs are permitted,
but
any surface
atom can hop to any nearest neighbor site at a rate
$D\exp(-nE_N/k_{\rm B}T)$, where
$D=(2k_{\rm B}T/h)\exp(-E_S /k_{\rm B}T)$ is the single
adatom migration rate, $E_N$
is an effective pair bond energy, and $n=0,1,2,3,4$ is the
number of lateral nearest neighbors before the hop occurs.
In the present work, the deposition and zero-strain
hopping rates are fixed
by the choices
$T=750\,$K,
$F=0.1 \; {\rm s^{-1}}$,
$E_S$=$1.3 \; {\rm eV}$,
and
$E_N$=$0.3 \; {\rm eV}$ so that
the dimensionless ratio
$D/F =  6 \cdot 10^5$. All
results
represent an average of at least 50
realizations on a
lattice of size 300$\times$300.

We
suppose \cite{APL}
that the principal effect of strain is to lower the barrier
to  detachment of atoms from  a strained pseudomorphic
island. In a mean field picture, one might choose the
barrier reduction to be identical for all atoms in an island
and equal to the average strain energy per atom of that island.
But since strain relief occurs primarily at the island edges we
instead adopt a scheme  whereby the strain-driven barrier
reduction depends on the local coordination number \cite{goofy}.
More precisely,
we make the replacement $E_N \rightarrow E_N - {\cal E}(\sqrt{s})$
for
the atoms of an island composed of $s$ atoms
where ${\cal E}(N)$
is the energy per atom
of a one-dimensional chain of $N$ harmonically coupled atoms in contact with
a rigid sinusoidal potential \cite{VdM}. This choice exploits
numerical results obtained from
a fully 2D version of
this model \cite{Snyman} which demonstrate that
the energy density of a square island is well
approximated by summing the energy density from two orthogonal
{\it non-interacting} chains of atoms.  Figure~1 shows a
log-log plot of the
strain-induced Arrhenius factor
$\exp[{\cal E}(s)/k_{\rm B}T]$ with
model parameters
chosen to reproduce the elastic and cohesive properties
of typical semiconductors \cite{Ratsch}.
 In practice, our numerical work makes use of the
the approximation to this quantity indicated by the
dashed straight line.

The main effect of strain is to promote the dissolution of large islands
by atom detachment \cite{BMB}. This is most
evident from
a plot of the number density of 2D islands of size $s$, $N_s(\Theta)$,
at fixed coverage  for
different values of misfit (Figure~2).
Note that the monomer population is not the only
beneficiary of the detachment process. Dimers and
other small islands form rapidly
but do not dissociate at an accelerated rate
because they are relatively unstrained.
The size distribution thus
both narrows and shifts toward smaller islands sizes.
More generally, we find that the
average island size  progressively decreases as misfit
increases at fixed coverage.
 We note in passing that, compared to homoepitaxy under otherwise
identical
conditions,
this observation implies that relatively {\it better} layer-by-layer growth
of strained material may be achieved so long as
kinetic barriers inhibit strain-driven
3D islanding or misfit dislocation generation.
 This  is so because
the probability to nucleate  next-layer islands onto the surface of
existing islands decreases very sharply as their size decreases \cite{Tersoff}.

For the case of {\it homoepitaxy}, it
is well established
 \cite{Evans,Stroscio,PRL,Bales,Family}
that the island size distribution at low coverage
 is  described
by
\be
N_s(\Theta) \sim {\Theta \over \langle s \rangle^2} \, g({s \over \langle s
\rangle})
\label{eq:ansatz}
\ee
where $g(x)$ is a scaling function and
$\langle s \rangle$
is the average island size.
To test (\ref{eq:ansatz}) for the case of {\it heteroepitaxy},
we plot $N_s \langle s \rangle^2 /\Theta$ versus
$s/ \langle s \rangle$ and ask whether the simulation
data at different coverages all collapse onto a single curve. That this
is indeed the case \cite{collapse} is illustrated in Figure~3 for
$5\%$ misfit and $10 \% \le \Theta \le 25 \%$.
Similar scaling is found for other
values of misfit
 so that, quite generally, the moments of the island
size distribution are given by
\be
M_n = \sum_s s^n N_s \sim \Theta \langle s
\rangle^{n-1} \int x^{n} g(x) dx.
\label{eq:moment1}
\ee

To make progress, we require the coverage dependence of $\langle s \rangle$.
The simulations reveal that this quantity is an increasing but not particularly
simple function of $\Theta$ for all values of misfit.
 But in the limited coverage
range noted above, it turns out that the power law
\be
\langle s \rangle \sim \Theta^z
\label{eq:average}
\ee
represents the data well. The misfit-dependence of the
exponent $z$ can be extracted directly
from (\ref{eq:average}) [square symbols in Figure 4(a)] or
 from the
coverage dependence of the density of all islands combined (dashed curves
in Figure 5) since, from
(\ref{eq:moment1}), the latter quantity takes the form
$N(\Theta) = M_0 \sim \Theta^{1-z}$ in the coverage range of interest.
The triangles in Figure 4(a) show that the two methods yield consistent
results.

To understand the observed monotonic decrease of $z$ toward zero it
is convenient to return to the full $N(\Theta)$ curves in Figure 5.
For our choice of deposition conditions,
$N(\Theta)$ grows
very rapidly independent of misfit below about $1.5 \%$ coverage.
 New island formation then slows
dramatically since the existing islands efficiently capture newly
deposited adatoms. But the strain energy per atom increases as the islands
grow and ejection of atoms from perimeter sites eventually ensues.
Since the ejection rate  increases as misfit
 increases, the
nucleation rate of new islands  from this source material increases
 similarly.
Comparison with the zero-strain case in Figure 5
reveals the efficacy of this process.

The coverage dependence of the number density of adatoms $N_1(\Theta)$
is shown in Figure~5 as well. The origin of the relative increase in
this quantity as a function of misfit is clear from the foregoing. More
interestingly, this quantity is seen to exhibit a power law variation
\be
N_1(\Theta) \sim \Theta^{-r}
\label{eq:monomer}
\ee
in the same coverage interval where $z$ was defined.
Figure 4(b) illustrates the misfit dependence of
the exponent $r$ extracted from the solid curves in Figure~5 \cite{caveat}.
The obvious question now arises: can a simple theory be constructed
that {\it predicts} the values for the exponents $z$ and $r$?
Previous simulation experience with aggregation-fragmentation phenomena
suggests that
rate equation theory may be adequate for this purpose \cite{previous}.
To this end, we
write an evolution equation for
the number density of each island species in the form \cite{BW}
\begin{eqnarray}
{dN_1 \over d\Theta} & = & 1 - K_1 N_1^2 - N_1 \sum_{s \geq 1} K_s N_s
+ {\gamma_2 N_2}
 + \sum_{s > 1} {\gamma_s N_s}
\label{eq:RE1} \\
{dN_s \over d\Theta} & = & N_1(K_{s-1}N_{s-1} - K_sN_s) - {\gamma_s N_s}
+ {\gamma_{s+1} N_{s+1}.}  \hskip 45pt (s > 1 )
\label{eq:RE2}
\end{eqnarray}
These mean-field  equations presume that only monomers are mobile and
that
islands grow and dissociate exclusively by the attachment and detachment of
single monomers.
The rate at which adatoms {\it attach}
to an island of size $s$ is assumed to take the form $K_s = K_0 s^p$
where, e.g., the exponent $p=1/2$ for the present case of 2D compact
islands \cite{BW}. Similarly,
the rate at which adatoms {\it detach} from an
island of size $s$ takes the form  $\gamma_s = \gamma_0 s^{\nu}$ for all
but the very smallest islands.
The exponent $\nu$ is deduced from
plots similar to Figure~1
to be $\nu \simeq 0.027 \langle n \rangle f^2$ where $\langle n \rangle$
is the average coordination number of the detaching species and
the lattice misfit $f$
is expressed in percent. Although $1 \le \langle n \rangle \le 2$,
we set $\langle n \rangle = 1$ in
what follows since the precise value of the coefficient of $f^2$
 anyway depends on the material
parameters used in the model calculation of ${\cal E}(\sqrt{s})$.

An exact equation of motion for the moments of $N_s$
follows immediately from
(\ref{eq:moment1}),(\ref{eq:RE1}) and (\ref{eq:RE2}):
\be
{d M_n \over d\Theta}  = \sum_s {((s+1)^n-s^n) (K_s N_1N_s + \gamma_s N_s)}
-\gamma_2 N_2 + K_1 N_1^2 + {dN_1 \over d\Theta}.
\label{eq:moment2}
\ee
But we are interested in a solution at
large times only so it is valid to neglect
low order moments and the last three terms
on the right hand side of
(\ref{eq:moment2}). This yields the approximate expression
\be
{d M_n \over d\Theta}  \simeq nK_0 N_1 M_{n+p-1} - n \gamma_0 M_{n+\nu-1}.
\label{eq:moment3}
\ee
When (\ref{eq:moment1}) is inserted into (\ref{eq:moment3}),
we obtain the following self-consistent solutions for the desired scaling
exponents
\be
r=z(p- \nu)
\hspace{0.8in} z = \left\{ \begin{array}{lccc}
(1 - \nu)^{-1}  & &  & \mbox{$0 \le \nu < p$} \\
& & &  \\
0 & & & \mbox{$\nu > p$}
\end{array}. \right.
\label{eq:relations}
\ee
Note that $z$ is indeterminate for the case $p$=$\nu$.
Otherwise, the values computed from (\ref{eq:relations})
are plotted as dashed curves in Figure~4
 for comparison with the Monte Carlo values.

It is obvious that
the prediction for
the monomer exponent $r$ show the same trend as
the simulation results while the exponent $z$ disagrees qualitatively.
Presumably,
the non-analytic
behavior of (\ref{eq:relations})
is an artifact of the simplifications required to derive (\ref{eq:moment3}).
But are these approximations also responsible
for the disagreement between the simulations and the rate equations
regarding the behavior of $z$?
To test this,
we solved
(\ref{eq:RE1}) and (\ref{eq:RE2}) numerically using rate parameters identical
to those used in the simulations. Scaling of the assumed
form does occur--but only at larger times (than found from the
simulations) where coalescence should be important. Be that as it may,
the qualitative behavior predicted by (\ref{eq:relations}) is confirmed
albeit with the discontinuities smoothed out (Figure~4).

Quantitatively, our numerical integration confirms a recent prediction by
Blackman and Marshall \cite{BM}
that $r$=$0$ and $z$=$1/(2-\nu)$
for the case $\nu$=$p$. Otherwise, our prediction that $r$=$z$=$0$  when
$\nu > p$
agrees with that of Ref. \cite{BM} but our
result that $z>1$ when $\nu < p$ is not consistent with
these authors conclusion that ``gelation takes place with the formation
of an infinite cluster'' in that regime. In any event, comparison
with the Monte Carlo results clearly impel us to
conclude that the failure of the rate equations to
reproduce the  exponent $z$ extracted from
the simulations is a real effect.

In the simulations (and in
reality) an atom that detaches from an island generally remains close to
that island and re-attaches to it with a high probability.
But in the rate equation treatment, an adatom that detaches from an
island becomes available for capture by all islands.
Moreover, since
larger islands capture more efficiently ($p > 0$), the average island
size will grow at an exaggerated rate
if monomers are in sufficient supply. Since increasing misfit precisely has the
effect of generating monomers (cf. Figure~5),
(\ref{eq:average}) implies that $z$ will
increase over its homoepitaxy value.
When $\nu > p$,
atoms are ejected from
larger islands at a higher rate than they are captured, many small
islands form,
and $z$
decreases precipitously. Eventually, the strain-induced reduction in
the detachment barrier exceeds the pair bond energy $E_N$ and all
islands disintegrate to yield adatoms as the only adsorbed species
($z \rightarrow 0$).

The present results
can be combined usefully with the fact that rate equation
predictions for homoepitaxy are satisfied
quantitatively only
for very large values of $D/F$ \cite{Evans,Family,Tang}. Both results suggest
that the theory is valid only when the number density of
monomers is sufficiently small
that the error associated with assigning equal
capture efficiency to all islands of the same size is negligible.
This observation might help guide  future research directed to
the formulation of an improved rate equation description
of the present problem.

Dimitri Vvedensky  kindly made available the
computing resources
used to perform the simulations. We thank
John
Blackman for a pre-publication copy of Ref. \cite{BM} and
Klaus Kern for an insightful remark on heteroepitaxial layer growth.
This work was supported by the U.S. Department of Energy,
a NATO travel grant and the Research Development Corporation of Japan.

\newpage
\noindent
{\bf Figure Captions}

\bigskip \bigskip
\noindent
{\bf Figure 1:} Log-log plot of the island size dependence of the
strain-induced Arrhenius factor $\exp[{\cal E}(\sqrt{s})/k_{\rm B}T]$.
The solid symbols denote values
obtained analytically as described in the text.
The straight
dashed line is the power law
approximation to it used in the simulations and the
rate equations.

\bigskip \noindent
{\bf Figure 2:}
Distribution of island sizes for
different values of misfit $f$ after 0.15 monolayers
have been deposited.
Results were obtained at $T=750$ K with $F=0.1 \; {\rm s^{-1}}$,
$E_S=1.3$ eV,
and $E_N=0.3$ eV.

\bigskip \noindent
{\bf Figure 3:}
Typical data collapse
for $5\%$ misfit. The data
shown represent coverages from $10\%$ to $25\%$.

\bigskip \noindent
{\bf Figure 4:}
Misfit dependence of the scaling exponents $r$ and $z$.
Results are shown
from
the Monte Carlo simulations at $0$\%, $3$\%, $4$\%, $4.7$\%, and $5$\% misfit
 for $10\% \le \Theta \le 25\%$ (squares and triangles),
an  analytic analysis of
the rate equations  (dashed lines), and a numerical solution of
the rate equations (circles). The parameters $\nu$ and the misfit $f$ are
related through $\nu \simeq 0.027f^2$.

\bigskip \noindent
{\bf Figure 5:}
Coverage dependence
of the number density of adatoms $N_1(\Theta)$ (solid curves) and the number
density of all other island species combined $N(\Theta)$ (dashed curves)
for different values
of misfit.


\begin{thebibliography}{99}
\bibitem{caveat}The error bars in Figure 4(b)
indicate that the square symbols  should
be regarded as upper bounds to the values of the exponent $r$. This is
so because
``direct hit'' events
where a deposited atom lands on top of or immediately adjacent to
an existing island are known to generate a downward bending
of the $N_1(\Theta)$ curve for the case of homoepitaxy \cite{PRL}.
\end{thebibliography}
\end{document}